\documentclass[a4paper]{iopart}\usepackage{iopams}  
\usepackage{graphicx}
\usepackage{epsfig}%nova org
\usepackage{eucal} %jxzj
\usepackage{amssymb}%nova org
\usepackage{amsfonts}%nova org
\usepackage{amsbsy}%jxzj uncommond

\begin{document}

\title{Self Interference of Single Electrodynamic Particle in Double Slit }
\author{J.X. Zheng-Johansson
%(December, 2007; January 20, 2008; Mar, 2010, April 3, (April 7)
}

\address{Institute of Fundamental Physics Research, 611 93 Nyk\"oping, Sweden\\
(April 28, 2010)
\footnote{Presentation at the 6th Int. Symp. Quantum Theory and Symmetries, University of Kentucky, Lexington, July 20-25, 2009.}
}
           %\ead{jxzj@iofpr.org}
%(April 14, 21)
%\address{
%1.  
%(December, 2007; January 20, 2008)
%} 
%
%\date{April, 2007}
%\date{December 17, 2006}
%\address{July, 2007}
%\address{November 7, 2007}
%\address{November 30, 2007}
\date{(April 28, 2010)}

\def\Pb{{\bf{P}}}
\def\ini{{\rm{i}}}
\def\fin{{\rm{f}}}
\def\hatu{\hat{u}_q}
\def\rbar{{\bar{r}}}

\def\thetabar{{\bar{\theta}}}
\def\La{L_\nu}
\def\Ja{J_\nu}
\def\ef{{\rm ef}}
\def\Ep{{P_1}}
\def\ioii{{\mbox{\normalsize${\frac{1}{2}}$}}}
\def\Isub{{\mbox{\tiny${I}$}}}
\def\IIsub{{\mbox{\tiny${II}$}}}
\def\ABsub{{\mbox{\tiny${AB}$}}}
\def\Asub{{\mbox{\tiny${A}$}}}
\def\ACsub{{\mbox{\tiny${AC}$}}}
\def\BPsub{{\mbox{\tiny${BP}$}}}
\def\Bsub{{\mbox{\tiny${B}$}}}
\def\Csub{{\mbox{\tiny${C}$}}}
\def\CPsub{{\mbox{\tiny${CP}$}}}
\def\Aa{A}
\def\Omegam{\mit{\Omega}}
\def\Omgsub{{\mbox{\tiny${\Omegam}$}}}
\def\Thetam{\mit{\Theta}}
\def\Thetasub{{\mbox{\tiny${\Thetam}$}}}
\def\Co{C_0}
\def\Wtil{{\tilde{{\mbox{\small${\Omegam}$}}}}}
%\def\Wtil{{\tilde{\Omegam}}}
%\maketitle %iaia
\def\as{p}

\def\psias{\psi}
\def\Phimas{\Phim}
\def\fas{f}
\def\ef{{\rm{ef}}}
\def\ub{{\bf u}}
\def\rei{{\rm{r}}}
\def\bw{b}
\def\mub{{\pmb{\mu}}}
\def\ub{{\bf{u}}}
\def\A{$A$ }
\def\B{$B$ }
\def\C{$C$ }
\def\P{$P$ }
\def\E{$E$ }
\def\Q{$Q$ }
\def\S{$S$ }

\def\Thetam{{\mit{\Theta}}}
\def\lsim{{< \atop \sim}}
\def\gsim{{> \atop \sim}}
\def\ev{\epsilon}
\def\Z{Z}
\def\vac{{\rm{vac}}}
\def\v{{\rm{v}}}
\def\Rsub{\mbox{\tiny${R}$}}
\def\ABsub{\mbox{\tiny${AB}$}}
\def\ACsub{\mbox{\tiny${AC}$}}
\def\OPsub{\mbox{\tiny${OP}$}}

\def\thetasub{\mbox{\tiny${\theta}$}}
\def\i{\mbox{\tiny{$I$}}}
\def\ii{\mbox{\tiny{$II$}}}
\def\iii{\mbox{\tiny{$III$}}}
\def\Ai{A$_{0}$}
\def\Bi{B$_{1}$}
\def\Bii{B$_{2}$}
\def\EMsup{{          {}_{\mbox{\tiny${em}$}}             }}
\def\EM{electromagnetic}
\def\lsub{\mbox{\tiny${l}$}}
\def\alphab{{\pmb{\alpha}}}
\def\sigmab{{\pmb{\sigma}}}
\def\Xim{{\cal X}}
\def\Pcal{{\cal P}}
\def\PsimR{\widetilde\Psim}

\def\Psima{\Psim}
\def\el{\mbox{${e}$}}
\def\elsub{\mbox{\tiny${e}$}}
\def\elmsub{\mbox{\tiny{${e}^{\minus}$}}}
\def\elpsub{\mbox{\tiny{${e}^{\p}$}}}

\def\gsub{\mbox{\tiny${\gamma}$}}

\def\Omegavel{\mathbin{{\mit\Omega}\mkern-13.mu^{_{\mbox{$-$}}}\hspace{-0.08cm}{}_d }}
\def\omegavel{\mathbin{{\mit\omega}\mkern-13.mu^{_{\mbox{$-$}}}\hspace{-0.08cm}{}_d }}
\def\Wvel{\Omegavel}
\def\wvel{\omegavel}
\def\nuvel{\mathbin{{\mit\nu}\mkern-13.mu^{_{\mbox{$-$}}}\hspace{-0.08cm}{}_d }}
\def\Nuvel{\mathbin{{\mit\Nu}\mkern-13.mu^{_{\mbox{$-$}}}\hspace{-0.08cm}{}_d }}

\def\Ci{1}
\def\betamt{{\bf{b}}}
\def\kbar{{\bar{k}}}
\def\kb{{\bf{k}}}
\def\Kb{{\bf{K}}}
\def\cb{{\bf{c}}}

\def\pb{{\bar{p}}}
\def\pbf{{\bf{p}}}
\def\Acal{{\cal{A}}}
\def\Bcal{{\cal{B}}}
\def\Ccal{{\cal{C}}}
\def\Ccalo{\Ccal_0}
\def\Vp{V}
\def\m{{\mbox{-}}}
\def\Ccal{{\cal{C}}}
\def\p{{\mbox{+}}}

\def\psipi{\psi_{\p}(1)}
\def\psipii{\psi_{\p}(2)}
\def\psimi{\psi_{\m}(1)}
\def\psimii{\psi_{\m}(2)}

\def\ai{\alpha(1)}
\def\aii{\alpha^{'}(2)}
\def\bi{\beta^{'}(1)}
\def\bii{\beta(2)}

\def\fa{f_r}
\def\fb{f_\ell}

\def\Ca{C_a}
\def\Cb{C_b}
\def\fbf{{\bf{f}}}
\def\Ocal{{\cal{O}}}
\def\psib{{\pmb{\psi}}}
\def\alphab{{\pmb{\alpha}}}
\def\sigmab{{\pmb{\sigma}}}

\def\Eb{{\bf E}}
\def\Bb{{\bf B}}
\def\ke{\kappa}
\def\nabb{{\pmb{\nabla}}}
\def\nablab{{\pmb{\nabla}}}
\def\vir{{\rm vir}}
\def\psitot{\psi}
\def\jb{{\bf{j}}}
\def\vel{v}
\def\velb{{\bf{v}}}
\def\velsub{\mbox{\tiny{$\vel$}}}

\def\Re{{\rm{Re}}}
\def\Imtr{I}
\def\citeUnif{5}
\def\App{}
\def\AppA{}
\def\AppB{}
\def\Qcal{{\mathcal{Q}}}
\def\Tcal{{\mathcal{T}}}
\def\Cross{Q}
\def\Ucal{{\cal{U}}}

\def\vphilim{f}
\def\ft{{\mathcal{B}}}
\def\vphibar{\mathbin{\varphi\mkern-12.5mu-}}

\def\vphi{\varphi}
\def\med{{\med}}

\def\Mcal{{\mathfrak{M}}}

\def\Sb{{\bf{S}}}
         \def\xia{{\mathcal{A}}}
\def\tha{\theta}

\def\nb{\bf{n}}
\def\zb{{\bf{z}}^0}
\def\phiv{\varphi}
\def\Lb{{\bf{L}}}
\def\velsub{{}_{\vel}}

\def\nablab{{\pmb{\nabla}}}
\def\velb{{\pmb{\vel}}}
\def\minus{\mbox{-}}
\def\m{\mbox{-}}

\def\Ab{{\bf{A}}_a}
\def\vel{\upsilon}
\def\Thm{\vartheta}
\def\lb{{\bf l}}
\def\vb{{\bf{v}}}

\def\Rb{{\bf R}}
\def\rb{{\bf r}}
\def\rbb{\as}
\def\rav{\bar{r}}
\def\pd{\partial}
\def\vphi{\varphi}

\def\empty{{\mbox{\tiny${\emptyset}$}}}

\def\psitot{\varphi}
\def\psiR{\widetilde{\psi}}
\def\psiL{\widetilde{\psi}^{{\rm vir}}}
\def\PhimR{\widetilde{ {\mit \Phi}}}
\def\PsimR{\widetilde{ {\mit \Psi}}}
\def\PsimL{{\widetilde{ {\mit \Psi}}}^{{\rm vir}}}
\def\a{\alpha}
\def\uav{\bar{u}}
\def\D{\Delta}
\def\th{\theta}
\def\r{{\mbox{\tiny${R}$}}}

\def\re{{\mbox{\tiny${R}$}}}
\def\Fmed{F_{{\rm a.med}}}
\def\med{{\rm med}}
\def\Lw{L_{\varphi}}
\def\Fb{{\bf{F}}}

\def\Efb{{\bf{E}}}
\def\Bfb{{\bf{B}}}
\def\Ac{ \varphi}
\def\Xsub{{\mbox{\tiny${X}$}}}
\def\Ysub{{\mbox{\tiny${Y}$}}}
\def\Zsub{{\mbox{\tiny${Z}$}}}

\def\Ksub{{\mbox{\tiny${K}$}}}
\def\W{{\mit \Omega}}
\def\Wd{\W_d{}}
\def\Nu{{\cal V}}
\def\Nud{\Nu_d{}}
\def\Eng{{\cal E}}
\def\eng{{\varepsilon}}
\def\Acuni{\Ac_{{\Ksub}^\dagsup}^{\dagsup}}
\def\unduni{\Ac_{{\Ksub}^\dagger}^{\dagsup}}
\def\Acauni{\Ac_{{\Ksub}^\ddagsup}^{\ddagsup}}
\def\Acunim{{\Ac_{{\Ksub}^\dagsup}^{\dagsup *}}}
\def\undunim{{\Ac_{{\Ksub}^\dagsup}^{\dagsup}}^*}
\def\Acaunim{{\Ac_{{\Ksub}^\ddagsup}^{\ddagsup *}}}
\def\pd{\partial}
\def\Ad{ {\mit \psi}}
\def\psim{ {\mit \psi}}
\def\Kd{K_d{}}
\def\Lam{{\mit \Lambda}}
\def\lam{\lambda}
\def\dagsup{{\mbox{\tiny${\dagger}$}}}
\def\ddagsup{{\mbox{\tiny${\ddagger}$}}}
\def\psimKdK{\psim_{\Ksub,\Kdsub}}
\def\w{\omega{}}
\def\wdlow{\omega_d }
\def\g{\gamma{}} 
         \def\Phim{{\mit \Phi}}
\def\Psim{{\mit \Psi}}
\def\arm{{\rm a}}
\def\brm{{\rm b}}
\def\crm{{\rm c}}
\def\drm{{\rm d}}
\def\erm{{\rm e}}
\def\frm{{\rm f}}
\def\grm{{\rm g}}
\def\hrm{{\rm h}}
\def\lf{\left}
\def\rt{\right}
\def\Kdsub{{\mbox{\tiny${K_d}$}}}
\def\psimkd{\psim_{\kdsub}}
\def\psimKd{\psim_{\Kdsub}}
\def\hquad{ \ \ } 
\def\Taum{{\mit \Gamma}}

%\maketitle %iop
%\tableofcontents

\begin{abstract}
It is by the long established fact in experiment and theory that  electromagnetic waves, here as one component of an IED particle, passing a double slit will undergo self inference each, producing at a detector plane fringed intensities. The wave generating point charge of a zero rest mass, as the other  component of the particle, is maintained a constant energy and speed by a repeated radiation reabsorption/reemission scheme, and in turn steered in  direction in its linear motion by the reflected radiation field, and will thereby travel to the detector along (one of) the optical path(s) of the waves leading to a bright interference fringe. We elucidate the process  formally based on first principles solutions for the IED particle and known principles for wave-matter interaction.
%\\{\scriptsize{Presentation at the 6th Int. Symp. on Quantum Theory and Symmetries, University of Kentucky, Lexington, July 20-25, 2009.}}

%\draft
%\keywords{ }

\end{abstract}

%\submitted %iop
%\maketitle %phyrev

\section{Introduction}
%\section{The double-slit challenge in quantum physics}
\label{Sec-intro}

Since first hypothesised by  L de Broglie in 1923\cite{deBroglie}, it has  been overall experimentally demonstrated that all matter particles, like the electrons and atoms,  classically pictured as  point entities,  manifest as also waves, referred to today as matter waves or quantum particles. These particles  are capable of  producing such wave effects as diffraction and interference entirely in the same way as the ordinary waves do. Overall experiments have subsequently corroborated that the matter waves formally are governed by  quantum mechanics, pillared with the Schr\"odinger equation, or alternatively the Heisenberg equation, and the Dirac equations. Up to the present however it has  remained as a basic open question as to what is waving with a matter wave ($\psi$). Regarding the nature of this wave, the current understanding has been  formally rested on an assumption that the complex form of it, $\psi(x,t)$, specifies the physical state of a particle assumed being spatially a point, with $|\psi(x,t)|^2$ describing the probability of finding the particle at location $x$ at time $t$. This in essence a "statistical point particle" picture has been successful in  accounting  for some of the basic properties of the quantum particles, in particular in predicting as eigen solutions the quantized dynamical quantities in direct agreement  with experiments.  This  picture however encounters difficulties in accounting for certain other important properties, most notably the superposition, diffraction and interference,  of the quantum-mechanical matter particles as well as "photons". 
The picture becomes paradoxical in the case of explaining the experimentally demonstrated  single particle interference (see e.g. the recent reviews in \cite{Mandel1999,Zeilinger1999,jxzjied}), most simply  in a double slit,  since  a point particle is not capable of entering two slits at the same time. The incompleteness with the current understandings of the matter waves  is commonly acknowledged. 

From the standpoint of the general physics of particles, the (single) particle  interference, or also the sometimes joined-in  quantum entanglement,  as we  see, is only one of a range of unsolved "mysteries"  in the realm of fundamental physics. 
By what scheme an electromagnetic  radiation is upon absorption or emission converted to or from particle's mass, by a portion or as a whole?
What is the origin of mass ($m$)? 
Why is $mc^2$, where $c$ is the velocity of light,  equal to the particle's total energy?   
Why do masses attract one another? 
And so forth. The eventual answers  to these questions in a coherent theoretical framework would demand a realistic particle scheme, or model. Such basic  properties of particles as  mass, total energy and wave function we know today  are  each dependent on  the particle's motion and these  in turn are convertible to and from electromagnetic radiation. This is to say that, the basic particle properties and their  relations are interrelated in a dynamical  context.  A particle scheme or model can not be an ultimate one which facilitates the predictions of some and yet not the overall of the  fundamental properties of particles.

Stemmed from general considerations as underlined above, the author developed in [\citeUnif a-l] based on overall experimental observations an internally electrodynamic  (IED) particle model, which briefly states that  {\it a single-charged material particle, like the electron, proton, etc., is constituted of (i) an oscillatory point  charge $q$ (as source) with a characteristic frequency  $\W$ and zero rest mass,  and (ii) the  electromagnetic waves generated by the charge.}  See  further a formal outline in Sec. \ref{sec-wavef}. The IED process is governed by  a minimal set of well established basic, or first principles  laws (for  a recent  review see e.g.  \cite{jxzjied}, [\citeUnif j]). A range of basic properties and relations for particles have become predictable in terms the first principle's solutions  for the IED process under corresponding conditions; solutions already obtained by the author include those reported in [\citeUnif a-l]. Based on first principles solutions for the IED particle and the known principles for  wave-matter interaction, we formally elucidate in this paper that  the IED particle will  naturally undergo self-interference in a double slit with  a scheme already built-in with the IED process.

\section{Wave equations, solutions for IED particle in three dimensions}
\label{sec-wavef}

Consider that a charge $q$ of a zero rest mass and having  an initial free-charge kinetic energy $\eng_q$ (hence a dynamical inertia) is inserted  at position $\Rb_q$ in the vacuum to which we attach  three Cartesian coordinates  $X,Y,Z$. Owing to  $\eng_q$, the charge is spontaneously  driven  into an oscillation (following a mechanical scheme elucidated e.g. in [\citeUnif j]) of a  characteristic frequency $\W$, of  displacement $\ub_q=u_q \hat{u_q}$ along  a $\hat{u_q}$ axis  about  $\Rb_q(X_q,Y_q,Z_q)$, and in addition,  a uniform translation at velocity $\vel$ in the $+X$-direction. In view of the general validity of the Maxwell's equations for the resulting waves below,  $u_q$ may be judged as relatively small and thus has the usual sinusoidal  solution (see e.g. [\citeUnif j]), assuming zero applied force,  $u_q(t)=A_q \sin( \W t+\a)$, with $A_q$ the amplitude and    $\a$ initial phase. The orientation of $\ub_q$, $\hatu$, may change  with time under the influence of  random environmental and/or applied fields. 

Suppose  for the present that  during a time $\delta t_1$, $\hatu$ is along the $Z$-axis; $u_q =Z_q'-Z_q$. Owing to its  accelerated motion $u_q$, the  charge  generates  electromagnetic waves of  electric fields $ \Eb^j(\rb,t)$'s and magnetic fields $\Bb^j(\rb,t)$'s  as propagated to position  $\rb(x,y,z)=\Rb(X,Y,Z)-\Rb_q$ in time $t=r/c$, $c$ being the velocity of light and $j$ indicating a source motion effect. The fields  are described by  the Maxwell's equations given for a specified $j$,  in regions excluding the charge $q$ and assuming no  other charge and current,   as 
$ \nablab \cdot \Eb^j =0,$
$ \nablab \cdot \Bb^j=0,   $
$\nablab \times \Bb^j=0 +\frac{1}{c^2}\pd_t \Eb^j,$
$\nablab \times \Eb^j =- \pd_t \Bb^j$. These  lead to  a wave equation for a $j$th component wave, denoting for brevity  by a  dimensionless wave displacement 
$\vphi^j \hat{\theta}=\Eb^j/E_q$ in the transverse $\hat{\theta}$ direction,
and a corresponding equation for the total wave $\psi=\sum_j \vphi^j$,   
$$\displaylines{
\refstepcounter{equation} \label{eq-waveq}
\hfill
 \pd_t^2 \vphi^j=c^2 \nabla \vphi^j \quad (a), 
\quad 
\pd_t^2 \psi=c^2 \nabla \psi  \quad (b)
\hfill (\ref{eq-waveq})}$$   
The travelling oscillatory charge $q$ and the  electromagnetic waves $E^j,B^j$'s, or  $\vphi^j$'s which are  continuously generated  and therefore in general  extend a wave train, make up our extensive IED particle,  with the prefix "extensive" unless for emphasis  omitted hereafter.

For $\vel <c$ (or $<<c$) holds ordinarily, there  thus ordinarily exists  a brief yet finite time interval $\delta t$ in which  the charge is essentially standing still, at $X_q$,  on the $X$ axis,  for which  $\a$ is a fixed value. So except for a source motion effect indicated by $j$, the solutions to the Maxwell's  equations are of the usual form $E^j=E_q \vphi^j$, where  $E_q\dot{=}\frac{\mu_0 q \w^{2} A_{q} }{4 \pi } $, with $\w\dot{=}\w^j$, and 
$$\displaylines{\refstepcounter{equation} \label{eq-vphitheta} \label{eq-vphithetaC}
\hfill
\vphi^\dagsup
=
\ioii\Ccal_0
 \sin(k_\as^\dagsup r -\w_\as^\dagsup t+\a ), \quad
\vphi^\ddagsup
=
-\ioii \Ccal_0 \sin(k_\as^\ddagsup r+ \w_\as^\ddagsup t -\a), 
\quad  
{\rm with}\ \Ccal_0= \Co \sin \theta /r,
\hfill(\ref{eq-vphitheta})
}$$
are two opposite travelling plane waves in the   $+r$- ($j=\dagger$) and $-r$- ($j=\ddagger$) directions along a certain radial path $\rb(r,\theta,\phi)$ in spherical polar coordinates $r,\theta, \phi$. In these, $k_\as^j= \g_\as^j K$,$ \w_\as^j= k_\as^j c =\g_\as^j \W$ are the Doppler displaced wavevectors and frequencies due to the source motion, with $\g_\as^\dagsup= \frac{1}{1-\vel_{\as}/c}, \g_\as^\ddagsup= \frac{1}{1+\vel_{\as}/c} $,  
 $\W=K c$, and  
$\vel_{\rbb}=\vel \sin \theta \cos \phi $ being the component  velocity of charge $q$ in the $+r$-direction;  $\Co$ is a normalization constant. 

Let the IED particle be confined as in later applications  between two massive, non absorbing vertical walls here separated at a distance $\La$ along  the $X$ axis.  Its waves  $\vphi^j$'s will travel up to  the walls, be  reflected,  reabsorbed  by the charge, partially and most  probably after a two-way reflection owing to detailed resonance, and then reemitted, repeatedly, thereby maintaining a constant total energy ($\eng_{tot}$) of the charge and wave system. If the charge has not emitted any radiation, then $\eng_{tot}=\eng_q$. If the charge has emitted its entire $\eng_q$ into electromagnetic waves of a mean  total  length $\Lw$ =$\sqrt{\Lw^\dagsup    \Lw^\ddagsup}$, of a total wave energy  $\eng (\equiv \eng_q)$,  then $\eng_{tot}=\eng $.  In representing a realistic extensive particle able to produce interference at the scale $\La$, $\Lw$ would wind about $\La$ in a large $\frac{J_\nu}{2}=\frac{\Lw}{2L_\nu}$ number of loops.  For this  extensive IED particle,  there will ordinarily be a portion $a_1$ of $\eng_{tot}$ conveyed by the charge and $a_2$ by the wave,   $a_1,a_2\le 1$, and $\eng_{tot}=a_1 \eng_q+\a_2 \eng$. For the IED particle in  stationary state mainly of our concern later, the rates of emission  $\a^{em}$, absorption  $\a^{ab}$, and transmission  $\a^{tr}$ by or   passing  $q$ need further be in equilibrium, $\a^{ab} =\a^{tr}+\a^{em}.$

At any point on a radial path  $r$ in $(0,\La)$  there will thus simultaneously prevail  the opposite travelling  $ \vphi^\dagsup$ and $\vphi^{\ddagsup}$, newly generated or reflected after round trips, of the IED particle. These superpose into a partial total wave $\psias=\vphi^\dagsup+\vphi^{\ddagsup}$, given after direct algebraic operations  in complex form as
$$\displaylines{\refstepcounter{equation} \label{eq-psia}
\hfill \psias(r,\theta,t)|_{k_{d\as}}
             % =\vphi^\dagsup+\vphi^{\ddagsup}
= \Phimas \fas, \quad \Phimas= e^{i[(K +  (\frac{\vel_{\rbb}}{c})k_{d \as} )r- (\frac{\vel_{\rbb}}{c})\w_\as t ]}, 
           %********************************************
           %= \cos[(K +  (\frac{\vel_{\rbb}}{c})k_{d \as} )r- (\frac{\vel_{\rbb}}{c})\w_\as t ], 
           %********************************************
\quad \fas
      =\Ccal_0 e^{i[k_{d\as}r - \w_\as  t+\a_0]}, 
           %********************************************
           %\Ccal_0 \sin [k_{d\as}r - \w_\as  t+\a_0], 
           %********************************************
 \hfill
\hfill  (\ref{eq-psia})
}$$
where $\a_0=\a-\frac{\pi}{2}$,
$\frac{1}{2}(k_\as^{\dagsup}-(-k_\as^{\ddagsup})) 
=K +\g_\as \frac{\vel_{\rbb}}{c} k_{d\as}
$,
$\frac{1}{2}(\w_\as^\dagsup-\w_\as^\ddagsup)
=\g_\as(\frac{\vel_{\rbb}}{c})\w_\as
\dot{=}\frac{\vel_{\rbb}}{c}\w_\as$,
$\frac{1}{2}(k_\as^{\dagsup}+(-k_\as^{\ddagsup}))
=\g_\as k_{d\as} \dot{=}k_{d\as}=\sqrt{(k_\as^\dagsup-K)(K-k_\as^\ddagsup)}$,
$\frac{1}{2}(\w_\as^\dagsup+\w_\as^\ddagsup)
=\g_\as \w_\as  
\dot{=}\w_\as= \sqrt{\w_\as^{\dagsup}\w_\as^{\ddagsup}} $, with the geometric means  being alternative expressions;
$\g_\as=\sqrt{\g_\as^\dagsup\g_\as^\dagsup}=1/\sqrt{1-\vel_{\rbb}^2/c^2}$ or  
$\g^2_\as =1+\g^2_\as (\vel_{\rbb}^2/c^2)$; and 
$$\displaylines{\refstepcounter{equation} \label{eq-kk3}
\hfill  k_{d\as}
= (\g_\as/\g) (\vel_{\rbb}/\vel) k_d,
\quad 
k_d =\g (\vel/c) K,  \quad 
 \w_\as= (\g_\as/\g) \w,
\quad
\w =\g \W,
\hfill (\ref{eq-kk3})
}$$
and $ \g=1/\sqrt{1-\vel^2/c^2}$. $\psias$ travels in the $+r$-direction at  a  phase velocity $W_{p\as}=\frac{\w_\as^\dagsup+\w_\as^\ddagsup }{k_\as^\dagsup+(-k_\as^\ddagsup)}=\frac{\w_\as}{k_{d\as}} $ and group velocity $W_{g\as}=\frac{\w_\as^\dagsup-\w_\as^\ddagsup }{k_\as^\dagsup-(-k_\as^\ddagsup)}= \frac{(\vel_{\rbb}/c)\w_\as}{K+ (\vel_{\rbb}/c)k_{d\as}}$, or   
$$\displaylines{\refstepcounter{equation} \label{eq-Wpas}
\hfill W_{p\as} 
= W_p/\sin \theta \cos \phi, \quad W_p=c^2/\vel,
\quad 
 W_{g\as}
\dot{=} \vel_{\rbb}
=W_g \sin \theta\cos\phi, 
\quad  W_g =\vel.
\hfill (\ref{eq-Wpas})
}$$
The $\psias$ due directly to the moving source  above is seen each $r,\theta$-dependent, or,  "polar spherical".

The electromagnetic waves have in terms of the total field $E_q\psi$  an energy density of the usual form  $\eng_0=\ev_0E_q^2|\psi|^2$, which  is transported at the speed of the $\vphi^j$ waves, $\w^\dagsup/k^{\dagsup}=\w^\ddagsup/k^{\ddagsup} =c$. The intensity of the total wave in relative terms is therefore  
$$\displaylines{\refstepcounter{equation} \label{eq-intena}
\hfill
I (r,\theta)=c\eng_{0}/c\ev_0E_q^2  = |\psi|^2  
={\Ccal_0} ^2.    
\hfill
(\ref{eq-intena})
}$$
The  $\eng_0$ integrated  over total solid angle $4\pi $ is  $\eng_{0x} =  \int_{4\pi} \eng_0 d \Wtil = \ev_0 E_q'{}^{2} |\psi_x| ^2$ where  $d \Wtil= \sin^2\theta d\theta d\phi$, $E_q'=\frac{2\sqrt{\pi}}{\sqrt{3}}E_q$,  and $\psi_x=\Co e^{i[k_{d}x - \w  t+\a_0]}$ is an apparent plane wave along the $X$-axis by which $\eng_{0x}$, being a constant, is effectively transported. Along the $X$-axis  $\theta=\frac{\pi}{2},\phi=0$,  the $\Ccal_0 (=\Co/x)$ and $I$ are a maximum each, $\psi=\psi_x/x$, and  the wave variables $k_d,\w$ describe  the dynamics of the mass center of the total  $\psi$ and hence the IED particle. The $X$-axis  therefore represents a symmetry axis of  $\psi$. 
The total wave energy across the full wave train  $\Lw(=\Ja \La)$ is  $\eng= \Ja\int^{\La}_0 \eng_{0x} d x  =\Lw \ev_0E_q'{}^2$ provided with  the normalisation $ \int^{\La}_0 |\psi_{x}|^2 d x=1 $ and thus $C_0=1/\sqrt{\La}$.  $p=\eng/c$ gives the total linear momentum.

Alternatively, employing the Planck energy equation for the $\psi$ of a frequency $\w$, and in turn  Newtonian kinetic energy equation for the wave train of $\psi$  which  travels rectilinearly at the finite velocity $c$ and accordingly has a finite inertial mass $m$, we may write down $\eng=\hbar \w=mc^2$; and $p(=\eng/c)=\hbar k=mc$ (see e.g. [\citeUnif a,c,e] for a detailed account). Making the expansion $\w(=\g\W) =\W +\wvel$  
where $ \wvel [= (1+\frac{3}{4}\frac{\vel^2}{c^2}+\ldots )\frac{1}{2}(\frac{\vel^2}{c^2}) \W]$, 
with  $\Eng=\lim_{\vel^2/c^2\rightarrow 0}\eng=\hbar \W$ and   $P=\lim_{\vel^2/c^2\rightarrow 0}p =\hbar K $, 
the differences $ \eng_\vel  (\dot{\equiv} \ioii m\vel^2) =\eng-\Eng  $ and $p_\vel (\equiv m \vel)=\sqrt{p^2-P^2}$ thus give  the kinetic energy and linear momentum of the IED particle. These further lead to  $M^2c^4 +p_\vel^2c^2 =\eng^2 $ and the corresponding  de Broglie relations
$$\displaylines{\refstepcounter{equation} \label{eq-deBroglieRela}
\hfill 
  \ioii m \vel^2= \hbar \wvel, \quad m \vel =\hbar k_{d}.
   \hfill (\ref{eq-deBroglieRela})
}$$
$\psi$ thus follows to represent a de Broglie phase wave 
                %in three dimensions 
and  reduces to a de Broglie wave $\Psim$ (Sec. \ref{Sec-ref})  at a thermal $k_d$ scale, each being polar-spherical.
More generally, at the scale $k_d$ equation  (\ref{eq-waveq}b)  reduces to a Schr\"odinger equation describing the kinetic motion of the particle[\citeUnif c,j]. 

From the above follows that   $K=Mc/\hbar >>k_d$  holds ordinarily, with $M=\lim_{\vel^2/c^2 \rightarrow 0}m$ the particle's rest mass.  The diffraction and interference effects to be dealt with later will be at the scale $k_d$, with  $\lam_d=2\pi/k_d$ the de Broglie wavelength, for which $ \Phimas \dot{=} 1,$ 
$
 \psi \dot{=} f=\Ccal_0 e^{i[k_{d\as}r - \w_\as  t+\a_0]}.
$

\section{Optical path of $\psi$ through a double slit. Interference fringes}
\label{sec-diff-dbw}

\begin{figure}[t]%hpb]
\begin{center} %{center}
\vspace{-.3cm}  %0.8
\includegraphics[width=1\textwidth]{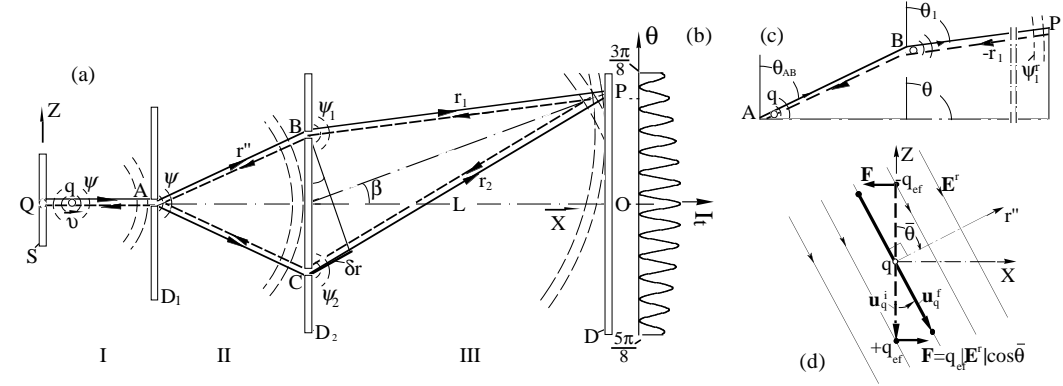}
\end{center} 
%\vspace{-.8cm}
\vspace{-1.cm}
 \caption{
(a) 
The  total electromagnetic  wave $\psi$ generated by travelling oscillatory charge $q$, with $\psi,q$ making up an IED particle, travels through a single slit $A$ and double slits $B,C$ to detector $D$. 
 (b)  Intensity $I_t$ (in units of $1/\Ccal_{\Bsub 0}$) of wave $\psi_t$ on $D$ calculated from (\ref{eq-inten}) for $b=1, \lam_d=0.5, \ell=4$, showing  two-slit interference fringes. 
 (c) Charge $q$ travels at velocity $\vel$ 
statistically along one 
 optical wave path $(Q)qABP$ through the double-slit up to $D$. (d) The effective dipole charges $+q_\ef,- q_\ef$ associated with the initial displacement  $\ub_q^{\ini}$ of $q$ along the  $Z$ direction, as at slit $A$, are acted on a torque $\Fb \times \ub_q$ by the reflected field $\Eb^{\rei}$ and turned into  aligned  with the  $\Eb^{\rei}$, in the $\thetabar$ direction, with  a final  $\ub_q^{\fin}$.
}
\label{fig1-2slits.eps}\vspace{-0.cm}
\vspace{.0cm}
\end{figure}

Suppose that the $\psias$ above, re-denoting now by $\psias(r',\theta',t')$, is first let strike on to an opaque screen $D_1$ with a narrow slit $A$ of width $b$, with $b\sim \lam_d$, at distance $x'$ from $q$, see Fig. \ref{fig1-2slits.eps}a. An  effectively plane wave, of a fraction  $g_\Asub$ of the integral  intensity ($I_{4\pi}=\int_{4\pi}I d \Wtil$),  only will enter the narrow slit, given by setting  $r'=x'$ and $\theta'=\frac{\pi}{2}$ in   (\ref{eq-psia}) as $ \psias{}(x',t')=\frac{C_\Asub}{x'}  e^{i (k_{d}x' - \w  t'+\a_{0})} 
$, where $C_\Asub=\sqrt{g_\Asub}\Co$. Along the $b$, the incident $ \psias{}(x',t')$ may be as in the usual approach thought  as if regenerated from many ($N$) small divisions of a width  $d s=b/N$ each, into a new wave  $\psi_s (r_s'',\theta_s'',t'') =\Ccal_0 e^{ [k_{d}r_s'' - \w  t''+\a_{\Asub}]}$ along a radial path $r_s''=r''+s \sin \beta''$ 
 joining a $ds$ at a vertical distance $s$ from the slit center $A$   to a point $P_2$ on an image plane $D_2$,   with $r''=AP_2$ and $ \beta''=\frac{\pi}{2}-\theta''$; $\a_\Asub=\a_0 + k_d x'_\Asub-\w t_{\Asub}$. The total diffracted de Broglie phase wave from $b$   arriving at $P_2$ at a later time $t''$, for $N$  being large and $x'$ assumed for illustration large, is given by 
$$\displaylines{
\refstepcounter{equation} \label{eq-psi-diff}
\hfill
 \psi (r'',\theta'',t'')|_{k_d}= \int^{\frac{\bw}{2}}_{\frac{-\bw}{2}}  \psi_s(r''_s,\theta_s'',t)  ds
=  \Ccal_\Asub  e^{i (k_d r'' - \w t''+\a_{\Asub})}, 
\quad  
\Ccal_\Asub (\theta'')=\frac{\bw \sqrt{g_\Asub}\Co\sin \eta''}{x' \eta''}
\hfill (\ref{eq-psi-diff})
}$$
 $\eta''= \frac{\pi  \bw \sin \beta''}{\lam_d}$. $\psi$ has  the phase  and group velocities $W_p$ and  $W_g$ as of $\psias(x',t')$, and an intensity as given after (\ref{eq-intena}) for  $\Ccal_\Asub$ here, $I(\theta'')=\Ccal_\Asub^2(\theta'')$.

Let the $D_2$  above be an opaque screen containing two narrow slits $B,C$ similarly of a width $\bw$ each, as in Fig. \ref{fig1-2slits.eps}a; and  the $\psi(r'',\theta'',t'')$ given by (\ref{eq-psi-diff}),  regenerated at $A$ at time $t''_0=0$, is incident on $D_2$. 
Two of its partial components  travelling along the equidistant radial paths $r''= AB$ and $AC$ at angles $\theta''=\theta_{\ABsub}$ and $\theta_{\ACsub}$ to  the $Z$ axis, and of a  fraction $g_{\Bsub}= g_{\Csub}$ of total intensity each,  $\psi(AB,\theta_{\ABsub},t'')$ and $\psi(AC,\theta_{\ACsub},t'')$ will at  a later time $t''$ ($=AB/c$) simultaneously arrive at   $B$ and $C$; here they have an identical  amplitude $\Ccal_\Asub $ and initial phase $\a_{\Asub}$. By the standard notion we know   that  each  electromagnetic wave $\vphi^j$ will on passing  the two slits undergo self interference. The formal conditions for this for the specific total,  de Broglie phase wave $\psi$ here are written down as follows.

The two slits $B,C$, as two new coherent sources, then regenerate at time $t_0=0$ the respective incident waves into two new spherical waves given at location $r$ time $t$ as 
$
\psi_{1}=\Ccal_1 e^{i(k_d r_1-\w t+\a_{\Bsub})}, 
\psi_{2}=\Ccal_2 e^{i(k_d r_2-\w t+\a_{\Bsub})}$ in region III 
similarly as given by (\ref{eq-psi-diff}), where $\Ccal_i =\Ccal_{\Bsub 0} \frac{\sin \eta_i}{\eta_i}$, with  $ i=1,2$,  $\Ccal_{\Bsub 0} =\frac{b \sqrt{g_\Bsub g_\Asub} C_0}{r'' }$, $\eta_{i}=\frac{\pi b \sin \beta_i}{\lam_d}$, and $\beta_i=\frac{\pi}{2}-\theta_i$. 
Let at distance $L$ apart in front of $D_2$ be placed a detector $D$,  and $P$ be a  point on $D$ as in Fig. \ref{fig1-2slits.eps}a. Two respective partial components  of $\psi_1$, $\psi_2$ travelling along paths $r_1=BP,r_2=CP$ at angles $\theta_{1},\theta_{2}$ to the $Z$-axis,   will at a later  time $t$ arrive at $P$, superposing linearly as $ \psi_{t} =\psi_1+\psi_2 $. Assuming as in usual applications $L>>BC$,  so for  the amplitudes,  the two paths $r_1$,$r_2$ can be  approximated as parallel to each other, of an average distance  $r=\ioii (r_1+r_2)$ and angle  $\theta =\ioii(\theta_1+\theta_2)$; thus $\Ccal_i \dot{=}\Ccal =\Ccal_{\Bsub 0}\sin \eta /\eta$, with   $ \eta=\pi b \sin \beta/\lam_d $, $\beta=\frac{\pi}{2}-\theta$. For the phase constants  the difference between the optical paths, $\delta r=r_1-r_2$, yet is nontrivial. 
Incorporating  the foregoing, 
the summation for $\psi_t$ is algebraically given as  
$$\displaylines{
\refstepcounter{equation} \label{eq-psipp}
  \hfill \psi_{t}       %=\psi_1+\psi_2 
           %****keep
           % = A_d \{ [\cos ([k_{d}r_1 - \w  t+\a_0) + \cos(k_{d}r_2 -\w  t+\a_0) ] \hfill \cr 
          %\hfill  \qquad\qquad +i[\sin ([k_{d}r_1 - \w  t+\a_0) + \sin(k_{d}r_2 -\w  t+\a_0) ] \}\hfill\cr\hfill \qquad \qquad
           %**keep
=\Ccal_t(\delta r) e^{i(k_{d}r -\w t +\a_\Bsub)},
\quad 
\Ccal_t(\delta r)
=2\Ccal\cos[\ioii k_{d} \delta r ].  \hfill (\ref{eq-psipp})            
 }$$
$\psi_t$, or the  field $\Eb_t=E_q \psi_t \hat{\theta}$ has a relative intensity  given similarly  as  (\ref{eq-intena})  as
$I_{t}=\frac{c \ev_0 E_q^2 |\psi_t|^2}{c \ev_0 E_q^2} $, or
$$\displaylines{\refstepcounter{equation} \label{eq-inten}
\hfill I_{t}=|\psi_t|^2 
=|\psi_1+\psi_2|^2
=\Ccal_t^2(\delta r) 
=4(|\psi_1|^2+|\psi_2|^2) \cos^2(\ioii k_d \delta r)
\hfill (\ref{eq-inten})
}$$
where $|\psi_i|^2=\frac{1}{2}\Ccal^2$, $i=1,2$. Now, 
$$\displaylines{\refstepcounter{equation} \label{eq-Brageq}\label{eq-Brageqb}
  \hfill {\rm if} \ \ \delta r = 2n\pi/k_d =n\lam_d, 
\quad
 n=0,1,2,\ldots,  \quad 
{\rm then}\ \
 |\Ccal_t (\delta r)| = 2\Ccal, \quad I_{t}= 4\Ccal^2   \hfill \qquad 
(\ref{eq-Brageqb})
}$$
are maxima, yielding bright fringes (peaks in Fig. \ref{fig1-2slits.eps}b) at distances $PO=L \tan \beta \dot{=} L \frac{\delta r }{\ell} =   n\lam_d L/\ell$ from $O$, where  $\ell=BC$. Or, if $\delta r=(n+\frac{1}{2})\lam_d$, then $\Ccal_t(\delta r)= I_t=0$
yielding  dark fringes at distances $PO=(n+\frac{1}{2})\lam_dL/\ell$ from $O$ on $D$.

\section{Reflection wave path}\label{Sec-ref}
A  wave $\psi$ (as $\psi_1$ or $\psi_2$ here) striking at the detector  $D$, assumed  massive and  not yet actualised a permanent absorption,  will be scattered or reflected back  from $D$, as $\psi^\rei$.  In determining the reflection paths, the following basic principles of electrodynamics and wave mechanics (a-c) apply: (a) The scattering of $\psi$ by a scatterer at  $P$ may be regarded as a temporary  absorption  of $\psi$ by  the scatterer, and then remission. (b) The amplitude of the absorption will be   large if $\psi$ and the scatterer are (nearly) in resonance,  or small or zero if otherwise. (c) The orientation of the symmetry  axis of $\psi^\rei$ is  determined by (c.1) the total linear momentum  $p'=\hbar k'=mc'$ of $\psi^\rei$  [after solutions to (\ref{eq-waveq})] and in addition, (c.2) the  path which requires the least action (the principle of least action). (d) For elucidating the mechanical scheme for the scattering of $\psi$ in vacuum here, a physical vacuum becomes indispensable which is proposed in  [\citeUnif a,g-h]
 based on overall experiments as composed of electrically polarisable  vacuuons.

\begin{figure}[htpb]
\vspace{-0.1cm}
\begin{center} %{center}
\includegraphics[width=0.95\textwidth]{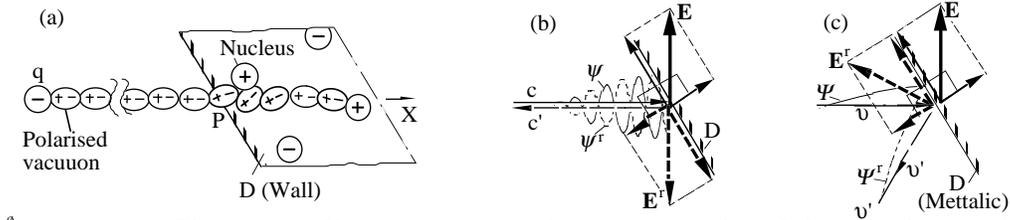}
\end{center} %{center}
\vspace{-0.8cm}
 \caption{(a)  Vacuuons  comprising the vacuum are
polarised about the charge $q$ of an IED particle, and  near 
$P$ at a massive wall $D$ in turn also by the mutually bound nuclear charges in the wall.
(b) Reciprocal reflection of  $\psi$ by $D$.
(c) Mirror reflection of $\Psim$ by  $D$.
}
\label{fig2-WavRefl.eps}
\vspace{-0.cm}
\end{figure}

For the $\psi$, vacuum and a massive  wall $D$ here we concretely 
find:
(\ref{Sec-ref}.1): With the $\w=\W +\wvel$ earlier, thus  
$\psi=\Thetam \Psim$, 
$ \Thetam =\Phim e^{-i\W t}$, 
$\Psim =\Ccal e^{i[ k_d r - \wvel t+\a_\Bsub]} 
$.
This  $\psi$ may  on the basis of (a)-(b) be scattered separately as  $\Thetam$ and $ \Psim$ by distinct scatterers.  The scatterer can only be a vacuuon (Fig.  \ref{fig2-WavRefl.eps}a-b)  for $\psi$ (or $\Thetam $ alone) whose frequency, $\w$ (or $\W$) of the scale of the particle's mass, is too high  to be (temporarily) absorbed by a material particle. For $\Psim$ of a  thermal frequency $\wvel$,  the scatterer    may be a material particle,  as illustrated in Fig.  \ref{fig2-WavRefl.eps}c for a free electron in the wall  of a metallic surface. (\ref{Sec-ref}.2): The vacuuons will be polarised by  the presence of $q$ or also by the passing of $\psi$, and additionally, when in their proximities by  the nuclear charges comprising the wall; they are thereby coupled with each other and in turn to the wall bulk  (Fig. \ref{fig2-WavRefl.eps}a). 
So when $\psi$ is incident onto the wall,  its field $\Eb$ acts on a vacuuon scatterer of an effective charge $q_p$) at the wall surface a force $\Fb(=q_p \Eb)$, opposite to which the massive wall acts back  a force  $\Fb^\rei=-\Fb$ (Newton's third law). This gives a reciprocal reflection of $\psi$, into $\psi^{\rei}$ of a linear momentum $mc' =-mc$ (Fig.  \ref{fig2-WavRefl.eps}b).
(\ref{Sec-ref}.3): 
For the $\psi_t$ of (\ref{eq-psipp}) striking on $D$, its reflected wave 
$\psi_t^\rei$, will be determined by (\ref{Sec-ref}.2) to have a  linear momentum in the reciprocal $-r$ direction and, while this being met,  will however practically choose to  retrace along the two separate  reciprocal paths $-r_1,-r_2$ that are already  (continuously) stretched by the incident waves and thus require a least action.
The optical reflection paths for $\psi_t$ from $D$ are therefore for large $L$ equal probably the $PBAqQ$ and $PCAqQ$ in Fig. \ref{fig1-2slits.eps}a (dashed lines). The reflection "wall" $S$ on the left  may  in real applications be a  particle beam injection  "gun"; its mid point $Q$ serves as a virtual source.

\section{Motion of $q$ through double slit
    }\label{Sec-q-path}

Provided that no permanent  emission of radiation occurs on its way, the point charge $q$ will travel across the double slit by the repeated radiation reabsorption/reemission scheme (Sec. \ref{sec-wavef}). The charge motion in region I is straightforward, and  through regions II-III  (Fig. \ref{fig1-2slits.eps}c) involves re-directions of its linear velocity by the reflected fields.  We elucidate the latter process and the physics involved below.

        %     %%%%%%%%%%%%%%%%%%%%%%%%%%%%%%%%%%
              %\footnote{Concretly,  for electron, $\vel=10^6$ m/s. Source -image plane distance is $L=1.5$ m as in \cite{Tonomura:1989}. The time interval $\delta T$ for an electron to move a distance $\delta X=1$ mm is $\delta T= \delta X/\vel= 10^{-9}$ s; in  $\delta T $ the distance traversed by $\psi$ at velocity $W_p=c^2/\vel$ is  $\Delta L= \delta T c^2/\vel = 90$ m which amounts to  30 loops about the total device. }
        %     %%%%%%%%%%%%%%%%%%%%%%%%%%%%%%%%%%

On arriving at  the slit $\Aa$ (Fig. \ref{fig1-2slits.eps}c), at initial time the charge  has a velocity $\vel$ and  symmetry  axis in the $+X$-direction, and   a $\ub_q^{\ini}$  and  radiation field $\Eb^{\ini}$ which are according to  (\ref{eq-intena}) invariably along the transverse, $Z$- direction. During a dwelling time $\delta t$ here it  will continuously generate radiation field and in turn be met by the reflected field  $\Eb^\rei$
from either side. By origin of electric field production in general,  the field here immediately due to  the $\ub_q^{\ini}= u_q \hatu'$ at $A$, $\Eb^{\ini}= \frac{E_q}{A_q} \ub_q^{\ini}$,  may be thought as produced between two effective dipole charges $+q_\ef,-q_\ef$ separated at a distance $\ub_q^{\ini}$   pivoted about $q$    (Fig. \ref{fig1-2slits.eps}d). So,  due to a $\Eb^\rei $ in a certain $\thetabar$- direction  the two dipole  charges will be acted   by two 
 forces $+\Fb,-\Fb=q_\ef  E^\rei \cos \thetabar $, hence a torque $\pmb{\tau}  = \ub_q^{\ini} \times \Fb =\pmb{\mu}_{q_\ef} \times \Eb^{\rei}$, with $\pmb{\mu}_{q_\ef}= q_\ef\bf{u}_q^{\ini}$, until after a relaxation  time $\delta t_1$,  its axis $\hatu{}' $ is
turned into $\ub^{\fin}=u_q \hatu''$  in  $\thetabar$-direction (Fig. \ref{fig1-2slits.eps}b).
(We may comprehend the microscopic physics as that,  the  vacuum is a dielectric[\citeUnif] and is produced with a polarization $\Pb_\v$ about $q$ for $q$ at  each fixed position, and an altered $\Pb_\v$ 
on deformation due to $\ub_q^{\ini}$, 
hence  the $+q_\ef,-q_\ef$, 
and in turn additionally due to the $\Eb^\rei $; the latter passes on no energy but a turn in the dipole axis  only.)
Accordingly, the charge will then be (re-)directed to travel  along the radial path $r''$   perpendicular to 
$\ub_q^{\fin}$.

In practice, $\psi^{\rei}$ is  reflected from all $ \rbar$-directions, and thus  with a  $\Eb^{\rei}$ in all $\thetabar$-directions in a statistical manner; $|\Eb^{\rei}|$ also varies  with time. The probability for $\ub_q^{\ini}$ to be turned to a $\ub_q^{\fin}$ parallel with a particular  $\Eb^{\rei}(\rbar,\thetabar,t'')$ is apparently in direct proportion to  the intensity of $\Eb^{\rei}$, $I(\thetabar)$ given in (\ref{eq-intena});  with $\psi$ of  (\ref{eq-psi-diff}) in it here we have $\Pcal(\thetabar)  =|\psi^{\rei}(0,\thetabar,t'')|^2
=\Ccal^2_\Asub(\thetabar) $. We suppose  in the following that, by a probability $ \Pcal(\theta_{\ABsub})$,  the charge $q$ at  $A$ has been re-directed to travel in the $AB$- direction and  $\hatu{}''$ along the transverse $\theta_{\ABsub}$-direction (Fig. \ref{fig1-2slits.eps}a),  and    at a later time $t''$ enters $B$ as in Fig. \ref{fig1-2slits.eps}d. 

Similarly, at $B$ the charge, with an initial  $\ub_q^{\ini}$ in the $\theta_{\ABsub}$ direction, is statistically subject to a torque due to a certain reflected field, which 
according to (\ref{Sec-ref}.4) is a projection of  a total $\Eb^\rei_t$ onto  $\thetabar =\theta_1$ or $\theta_2$ perpendicular to    $\rbar=-r_1$ or $-r_2$.
Particularly here, $\Eb^\rei_t$, as the incident $\Eb_t (=E_q\psi_t \hat{\theta})$, has  peaked intensities $I^\rei_t(\thetabar)=I_t(\theta)$ given by (\ref{eq-Brageq}); and its projection on $\theta_i$, with $i=1,2$, 
is $I^\rei_1(\thetabar{}_i)=\frac{1}{2}I^\rei_t(\thetabar)$ following from the effectively  symmetric geometry  of $P$ relative to  $B,C $ for large $L$ in region III, and $I(\theta''_{\ABsub})=I (\theta''_{\ACsub})$ in region II.
The probability for $\hatu{}''$ to be turned to $\hatu{}$ parallel with the $\thetabar_{i}=\theta_i$- direction perpendicular to $\rbar_i=-r_i$, $i=1$ or $2$, in an individual measurement 
is thus given by
$$\displaylines{
\refstepcounter{equation} \label{eq-probp} \label{eq-probpav}
\hfill \Pcal_i (\theta_i) =\ioii \Pcal_t(\theta)
=\ioii (|\psi_1|^2+|\psi_2|^2)
4 \cos^2(\ioii k_d \delta r)= \ioii |\psi_1+\psi_2|^2,   \quad i=1,2.
 \hfill (\ref{eq-probp})
               %\cr   \hfill  \ioii<\Pcal_1+ \Pcal_2> = \ioii \Pcal_t (\theta), =|\psi_1+\psi_2|^2,  \quad  \delta r= n\lam_d, \quad n=0,1,2,\ldots  \hfill(\ref{eq-probpav}b)
}$$
(\ref{eq-probpav}) implies that, in each individual measurement the point charge $q$  has definitely, equally probably travelled through either slit $B$ or $C$  before striking on $D$. The detection signal produced directly by the striking of $q$ however does not inform  which slit is passed by $q$. 
A determination of which slit can be prevented, not in principle but in common practice. If for example one attempts to determine which slit by blocking one of the two, say $C$, then the intensities of the incident and reflected waves and of the so navigated $q$  will each  be of a pattern of single-slit diffraction. This feature has been well  demonstrated in  matter-wave double-slit experiments.

A detection signal at $D$ may be directly due to the impingement of $q$ or $\psi_t$ (in the latter case the slow moving $q$ may not have entered $D$),  of an  identical intensity $I_t$ given by (\ref{eq-inten}) or (\ref{eq-probpav}). Either will  inform the self interference of  the IED particle. From either (\ref{eq-inten}) or (\ref{eq-probpav}),  the two probabilities $|\psi_1|^2,|\psi_2|^2$ for  the single particle passing two slits  add up to the  total $I_t$   in the manner of  $|\psi_1+\psi_2|^2$, and not $|\psi_1|^2 +|\psi_2|^2$. This  predicts an important characteristics of quantum-mechanical  probability as  has been demonstrated by overall matter-wave interference experiments.
      %**** (see e.g. \cite{wlu.caGhose}), ****
        %
This way how the intensities add up is completely the same  as how the intensities of an elastic  or light wave add up; and  we have factually just obtained (\ref{eq-inten}) naturally for the IED particle for its constituent electromagnetic waves $\vphi^j$'s, or "light" rays of far-above ultra violet frequencies.

An incident IED particle with a  low enough energy  will knock  a point-like absorber in $D$ into excitation of one energy quantum $\hbar \wvel$ at a time, producing a detection signal appearing as a spatial point on the $D$ plane. When this is directly due to a (thermal) absorption of the extensive ($\Psim$ off from) $\psi$, we can  think of the wave as (conveying) an energy flux  and its each wave front as an  equi-energy surface $S$. An energy will first of all be transported to  $P$  longitudinally 
through the propagation of $\psi$ at the speed $c$.  In the meantime, a sudden drop of energy at  $P$ (upon onset of  excitation)  creates an energy potential against other points on the $S$  intersecting with  $P$,  driving therefore a lateral energy flow toward $P$ at a speed which may be expected to be high since no time is wasted in undulation.

The author expresses thanks to Professor H.-D. Doebner for his invaluable discussions at the 6th Int Symp Quantum Theory and Symmetries (QTS 6, Univ Kentucky), extending  over a number of previous ones, with the author  over various  key aspects, benefiting especially clearer presentations, of the IED representation and solutions,  to the QTS 6 organisers for arranging the presentation of this work at the QTS 6 meeting, to Scientist P.-I. Johansson for his continued moral and private funding support of the research,  to Scientists P.-I. Johansson and  H. Rundlof  for first-hand knowledge  discussions on pair annihilation and neutron diffraction measurement related subtle issues, and to Professor P Dini for coordinating two previous abstract communications of this work at ICQNM09,10.
During the QTS 6 meeting  the author especially also enjoyed the kind reception of Professor A. Shapere, QTS6 organising chairman, and
intellectual conversations with Professors M. Berry,  A. Shapere, F.  Verstraete,  P.  Devereus, 
J. Towe, 
V. Dobrev, 
V. Borowiec, 
and others.

\end{document}